\documentclass[twocolumn,aps,prb,superscriptaddress,showpacs]{revtex4}

\usepackage{graphicx,epstopdf,color}
\usepackage{amsfonts}
\usepackage{amsmath,amssymb,mathrsfs}
\usepackage[varg]{txfonts}
\usepackage{bm}

\newcommand{\bs}[1]{\boldsymbol{#1}}

\newcommand{\ket}[1]{\left|#1\right\rangle}

\newcommand{\nn}{\nonumber}
\def\ie{\emph{i.e.},\ }

\def\ea{\emph{et~al.}}

\begin{document}
 \title{Spinon confinement and the Haldane gap in SU($\bs{n}$) spin chains}

\author{Stephan Rachel}
\affiliation{Institut f\"ur 
Theorie der Kondensierten Materie, 
Karlsruhe Institute of Technology, 76128 Karlsruhe, Germany}
\author{Ronny Thomale}
\affiliation{Institut f\"ur 
Theorie der Kondensierten Materie, 
Karlsruhe Institute of Technology, 76128 Karlsruhe, Germany}
\author{Max F\"uhringer}
\affiliation{Institut f\"ur 
Theorie der Kondensierten Materie, 
Karlsruhe Institute of Technology, 76128 Karlsruhe, Germany}
\author{Peter Schmitteckert}
\affiliation{Institut f\"ur Nanotechnologie, Karlsruhe Institute 
of Technology, 76344 Eggenstein-Leopoldshafen, Germany}
\author{Martin Greiter}
\affiliation{Institut f\"ur 
Theorie der Kondensierten Materie, 
Karlsruhe Institute of Technology, 76128 Karlsruhe, Germany}

 \pagestyle{plain}
\begin{abstract}
  We use extensive DMRG calculations to show that a classification of
  SU($n$) spin chains with regard to the existence of spinon
  confinement and hence a Haldane gap obtained previously for valence
  bond solid models applies to SU($n$) Heisenberg chains as well.  In
  particular, we observe spinon confinement due to a next--nearest
  neighbor interaction in the SU(4) representation $\bs{10}$ chain.
\end{abstract}


\maketitle


{\it Introduction.---}The properties of quantum spin chains have been
a vital area of research in condensed matter physics.  Starting with
Bethe's solution of the spin\,$\frac{1}{2}$ Heisenberg model (HM) in
1931\,\cite{bethe31zp205}, the field quickly evolved and significantly
influenced many other areas of physics.  In accordance to previous
findings by Andrei and Lowenstein\,\cite{andrei-79prl1698}, Faddeev
and Takhtajan\,\cite{faddeev-81pla375} observed in 1981 by
consideration of Bethe Ansatz solutions that the elementary
excitations of spin chains are spinons carrying spin $1/2$.  From the
identification of the O(3) nonlinear sigma model as the low-energy
field theory of antiferromagnetic SU(2) spin chains, Haldane
conjectured in 1983 that chains with integer spin possess a gap in the
magnetic excitation spectrum, while a topological term renders
half-integer spin chains gapless\,\cite{hc}.  This, and in particular
the gap in the magnetic energy spectrum for integer chains, was
confirmed by experiment\,\cite{buyers-86prl371}.
An elegant paradigm of the gapped spin\,$1$ chain in terms of a
valence bond solid (VBS) model was given by Affleck, Kennedy, Lieb,
and Tasaki (AKLT)\,\cite{affleck-87prl799}.

Haldane's classification\,\cite{hc,affleck89jpcm3047} applies only to
SU(2) spin systems.  In the field of ultracold atoms and optical
lattices, however, experimental realizations of SU($n$) spin systems
may be possible in the near future.  In particular, it has been
proposed very recently that SU($n$) antiferromagnets with $n$ up to 10
may be realized with ultracold alkaline-earth atoms without any need
to fine-tune any of the interaction parameters~\cite{gorshkov}.  This
perspective makes a general classification of the magnetic spectra of
antiferromagnetic SU($n$) spin chains highly desirable.

It is, however, not immediately clear how to address the question
which SU($n$) spin chains generally possess a Haldane type gap, as
Haldane's original work cannot directly be generalized to SU($n$);
previous attempts were limited to special
cases\,\cite{affleck-86lmp57,affleck88npb582}.  One possible route is
to interpret the Haldane gap as the zero point energy of the
oscillator describing the relative motion of pairwise confined
spinons\,\cite{greiter-zeropoint}, and to employ simple paradigms to
determine whether spinons in a spin chain with spins transforming
according to a given representation are confined or not.  Following
this line of reasoning, two of us\,\cite{greiter-07prb184441} recently
formulated a variety of VBS models for SU($n$) spin chains, and
investigated which models exhibit spinon confinement and hence a
Haldane gap.  This led to a classification of SU($n$) spin chains into
three categories, as reviewed below.  The conjecture which motivated
this work was that the ``gap/no gap'' classification found for the VBS
models would apply to general SU($n$) spin chain models, and in
particular to the SU($n$) HMs.

In this paper, we explicitly confirm this conjecture using extensive
Density Matrix Renormalization Group (DMRG) calculations.  In particular, we observe confinement of spinons
and hence a change of the universality class triggered by a
next--nearest neighbor interaction in the simplest example of the
third category, the SU(4) rep.\ {\bf 10} model.


{\it Spinon confinement in VBS states.---}%
Arovas \ea\,\cite{arovas-88prl531} pointed out that the spin\,$1$ AKLT model
$H_{\rm AKLT}=\sum_i P_{\bs{2}}(i,i+1)$ 
is a good starting point to reach the Heisenberg point $H_{\rm
  HM}=\sum_i\bs{S}_i\bs{S}_{i+1}=
3\sum_i(P_{\bs{2}}(i,i\!+\!1)+1/3P_{\bs{1}}(i,i\!+\!1)-2/3)$
perturbatively. Here $P_{\bs{R}}(i,\dots,i\!+\!m)$ is a projector of
the total spin of $m+1$ neighboring sites onto representation $\bs{R}$
of SU(2).  Both models show an excitation gap according to Haldane's
conjecture.  For the AKLT model, exact
calculations\,\cite{arovas-88prl531} showed that the static spin--spin
correlations decay exponentially, being the signature of an excitation
gap (see also the second line in Tab.\,\ref{tab:corr}).  The AKLT
model contains a characteristic length scale associated with the
correlation decay, which may only change quantitatively, but not
qualitatively upon moving to the HM.  In the following, we assume that
also beyond SU(2), the existence of a length scale at a VBS point
persists as one moves to the Heisenberg point.  This links the gap
property of a VBS model with the HM for the same spin representation.

\begin{figure}[t]
\centering
\includegraphics[scale=1.0]{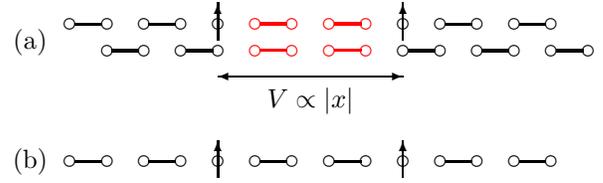}
\caption{(Color online) (a) Spinon confinement in the AKLT model.  The
  potential is proportional to the distance $|x|$ between the spinons
  and allows an interpretation of the Haldane gap as the zero point
  energy of a linear oscillator.  (b) Spinons in the spin 1/2 VBS
  state are domain walls between the two ground states and hence
  free.}
\label{fig:ho}
\end{figure}

It is easy to illustrate how the Haldane gap can be traced to spinon
confinement in the AKLT state.  It can be written as a product of
local spin singlets consisting of two "virtual" spins\,$1/2$ placed on
adjacent sites.  Projecting onto the symmetric subspace on each site
gives the spin 1 representation:
\begin{equation}
\ket{\Psi_{\rm AKLT}}=|
\setlength{\unitlength}{1pt}
\begin{picture}(128,10)(-7,2)
\linethickness{0.8pt}
\multiput(0,8)(14,0){9}{\circle{4}}
\multiput(0,2)(14,0){9}{\circle{4}}
\thicklines
\multiput(2,8)(28,0){4}{\line(1,0){10}}
\multiput(16,2)(28,0){4}{\line(1,0){10}}
\thinlines
\put(37,-3){\dashbox(10,16)}
\end{picture}
\rangle
\nn\label{akltstate}
\end{equation} 
Here, each circle denotes a ``virtual'' spin\,$1/2$ and each line a
singlet bond.  The dashed box illustrates exemplarily a lattice
site. The AKLT Hamiltonian is just the sum over the projectors onto
the subspace with total spin\,$2$ of adjacent sites.  For two
neighboring sites of $\Psi_{\rm AKLT}$, a spin singlet and two
individual ``virtual'' spins $1/2$ are present, \ie
$\bs{0}\otimes\bs{\frac{1}{2}}\otimes\bs{\frac{1}{2}}=\bs{0}\oplus\bs{1}$.
The tensor decomposition of two spin 1 representations, however,
additionally contains a spin\,$2$ subspace, \ie
$\bs{1}\otimes\bs{1}=\bs{0}\oplus\bs{1}\oplus\bs{2}$.  Hence, it
follows that $\Psi_{\rm AKLT}$ is annihilated by $H_{\rm AKLT}$,
whereas all other states will be lifted to finite energy.  To create a
pair of spinons, one breaks a singlet bond (see Fig.\,\ref{fig:ho}a).
The region between the two spinons yields an energy penalty for
$H_{\rm AKLT}$, or stated differently, the two spinons feel a
confining potential $V(x)=F|x|$.  As the Hamiltonian of the relative
motion of the two spinons consists of a kinetic energy and this linear
confining potential, we obtain a linear oscillator.  We interpret the
zero point energy of this oscillator as the Haldane gap.

For half--integral spin chains, the corresponding VBS is the
Majumdar--Ghosh (MG) model\,\cite{majumdar-69jmp1399} (see also the
first line in Tab.\,\ref{tab:corr}).  In our terminology both
spontaneously $n$-merized (like MG) and translational invariant states
(like AKLT) are VBS states.
Its ground state consists of local singlet bonds placed on adjacent
sites.  In contrast to the AKLT model, the MG ground state is
two--fold degenerate and invariant only under translations by two
lattice sites.  As in the AKLT model, elementary excitations are
created by breaking a singlet bond.  The two emerging spinons can
serve as domain walls interpolating between the two ground states (see
Fig.\,\ref{fig:ho}b), \ie the spinons do not feel any confining
potential. Accordingly, the spin--spin correlations decay abruptly,
\ie $\langle\bs{S}_i\bs{S}_{i+2}\rangle=0$, and no associated length
scale describing an exponential decay emerges. Moving from the MG
point to the isotropic HM, one passes a phase transition from the
dimer phase into a critical spin liquid phase~\cite{okamoto-92pla433}.
The deconfined spinons of the MG model, however, remain deconfined
when moving to the generic HM and give rise to the observable
two--spinon continuum\,\cite{tennant-95prb13368}.  Applied to
arbitrary integer and half integer representations, this line of
argument reproduces Haldane's original conjecture for SU(2) spin
chains. This interpretation is consistent with earlier attempts to
describe the spin 1 elementary excitation of related models as
confined solitons\,\cite{soliton-confinement} or two--spinon composite
particles\,\cite{brehmert-97jpcm1103}.


{\it Previous conjecture.---}So far, we reproduced Haldane's
conjecture for SU(2) in terms of spinon confinement.  We now review
the general classification of the magnetic excitation spectra for
SU($n$) chains motivated by the VBS states introduced
previously\,\cite{greiter-07prb184441} .  This classification depends
only on the degree $n$ of SU($n$) and the number of boxes $\lambda$ in
the Young tableau (YT) associated with the respective spin
representation (see also Tab.~\ref{tab:corr}):
(I) If $\lambda$ and $n$ have no common divisor, the model will
  support free spinon excitations and not exhibit a Haldane gap.
(II) If $\lambda$ is divisible by $n$, the model will exhibit spinon
  confinement and hence a Haldane gap.
(III) If $\lambda$ and $n$ have a common divisor different from $n$,
  the presence of the confining potential will depend on the range of
  interactions.  If $q$ is the largest common divisor of $\lambda$ and
  $n$, interactions ranging to the $(n/q)$-th neighbor are required for
  spinon confinement.  

  For SU(2), the classification is identical to Haldane's, as the
  third category becomes accessible only for $n\ge 4$.  If this
  classification applies to the HMs as well, as we advocate here, we
  will find gapless spectra, unique ground states, and algebraic
  correlations for the first category.  For the second category, we
  will find gapped spectra and exponentially decaying correlations, as
  it is possible to write the HM as VBS model plus a small
  perturbation (similar as for the spin 1
  case\,\cite{arovas-88prl531}). For the SU(4) rep.\,$\bs{10}$ chain
  belonging to the third category, the gapless spectrum of a
  nearest-neighbor HM should acquire a gap as a next-nearest neighbor
  interaction term is added.

\begin{table}[t]
   \centering
   \begin{tabular}[c]{lccc} 
\hline\hline
\raisebox{-5pt}[5pt]{~~~Spin rep.~~~~~~}  & \raisebox{-5pt}[5pt]{~~~~~~YT~~~~~~} 
& \raisebox{-5pt}[5pt]{~~~~~~~~$\xi$~~~~~~~~} & \raisebox{-5pt}[5pt]{~~~~~C~~~}\\
  &  \\
\hline
SU(2), ${S\!=\!\frac{1}{2}}$ &
\setlength{\unitlength}{6pt}
\begin{picture}(1,1)(0,0)
\linethickness{0.3pt}
\multiput(0,0)(1,0){2}{\line(0,1){1}}
\multiput(0,0)(0,1){2}{\line(1,0){1}}
\end{picture}
& -- & I \\[5pt]
SU(2), ${S\!=\!1}$ &
\setlength{\unitlength}{6pt}
\begin{picture}(2,1)(0,0.5)
\linethickness{0.3pt}
\multiput(0,0)(1,0){3}{\line(0,1){1}}
\multiput(0,0)(0,1){2}{\line(1,0){2}}
\end{picture}
& $\frac{1}{\ln{3}}$ & II \\[5pt]
SU(3), ${\bs{6}}$ &
\setlength{\unitlength}{6pt}
\begin{picture}(2,1)(0,0.5)
\linethickness{0.3pt}
\multiput(0,0)(1,0){3}{\line(0,1){1}}
\multiput(0,0)(0,1){2}{\line(1,0){2}}
\end{picture}
& 
& I\\[5pt]
SU(3), ${\bs{8}}$ &
\setlength{\unitlength}{6pt}
\begin{picture}(2,1)(0,1.)
\linethickness{0.3pt}
\multiput(0,0)(1,0){2}{\line(0,1){2}}
\multiput(2,1)(1,0){1}{\line(0,1){1}}
\multiput(0,1)(0,1){2}{\line(1,0){2}}
\multiput(0,0)(0,1){1}{\line(1,0){1}}
\end{picture}
& $\frac{1}{\ln{8}}$ & 
II \\[6pt]
SU(4), ${\bs{10}}$ &
\setlength{\unitlength}{6pt}
\begin{picture}(2,1)(0,0.5)
\linethickness{0.3pt}
\multiput(0,0)(1,0){3}{\line(0,1){1}}
\multiput(0,0)(0,1){2}{\line(1,0){2}}
\end{picture}
& $\frac{2}{\ln{7}}$ & III\\[5pt]
\hline\hline
   \end{tabular}
   \caption{Classification for spin--spin correlations of certain 
     SU($n$) VBS states as introduced in 
     Ref.\,\onlinecite{greiter-07prb184441}. 
     Listed are the spin representation, the 
     corresponding Young Tableau (YT), the spin-spin correlation length 
     $\xi$\,\cite{GRSunpub}, 
     and the category C w.r.t.\ our conjecture.  The SU(4) rep.\,$\bs{10}$ 
     VBS is the simplest model of category III. The correlations 
     for the MG and the rep.\,$\bs{6}$ model decay abruptly and no 
     correlation length can be defined.}
   \label{tab:corr}
\end{table}
In Tab.~\ref{tab:corr}, we further list the exact values for the
static spin--spin correlation length~\cite{GRSunpub,TMM,TMM2}. 
We expect that Heisenberg chains for all
representations without a well--defined VBS correlation length $\xi$
(like line 1 and 3 in Tab.~\ref{tab:corr}) will be gapless.  For
representations with a well--defined correlation length $\xi$, we have
to distinguish a "standard" gapped category II and the non-trivial
category III, which we will treat separately below.


{\it Verification of the conjecture.---}%
To verify our predictions for the SU($n$) HMs, we performed
extensive DMRG\,\cite{white92prl2863,schollwoeck05rmp259,noack-proc,hallberg06ap477} studies for all SU($n$) representations up to the
ten--dimensional representations of SU(3) and SU(4). Note that we use the ``Abelian'' quantum numbers of SU($n$), as the ''non-Abelian version'' of DMRG\,\cite{mcculloch-nonabelian} is not 
really convenient for SU(3) and SU(4) representations. In order to treat the large block sizes of our DMRG calculation we use a Posix thread
parallelized DMRG code\,\cite{schmitteckert-xyz}.
In our list of representations, we find gapless models
such as the fundamental representations of SU(3) ($\lambda=1$ in YT
box notation), SU(4) ($\lambda=1$), and the six--dimensional
representation of SU(3) ($\lambda=2$).  The ten--dimensional
representation of SU(3) (\setlength{\unitlength}{6pt}
\begin{picture}(3,1)(0.15,0.1) \linethickness{0.3pt}
  \multiput(0,0)(1,0){4}{\line(0,1){1}}
  \multiput(0,0)(0,1){2}{\line(1,0){3}}
\end{picture}, 
$\lambda=3$) provides an example for a gapped model.  As the local
basis dimension can become very large, we are not always able to
extrapolate to the thermodynamic limit and read off the gap from
there.  Instead, we use the entanglement entropy
$S_\alpha=-\text{Tr}[\rho_\alpha \log \rho_\alpha]$, where
$\rho_\alpha$ is the reduced density matrix in which all the degrees
of freedom on sites $i>\alpha$ are traced out. This way, there is no
need to extrapolate.  In particular, for critical spin models
associated with conformal field theories (CFT), the entanglement
entropy is given by\,\cite{calabrese-04jsmp06002}
\begin{equation}
  \label{ee-formula}
  S_{\alpha,L}=\frac{c}{3}\log{\left[ \left(\frac{L}{\pi}\right)
\sin{\left(\frac{\pi\alpha}{L} \right)} \right]}+c_1\ ,
\end{equation}
where $L$ is the chain length, $c_1$ a non-universal constant, and $c$
the central charge of the associated CFT.  In case of Heisenberg--like
SU($n$) spin chains, the CFT is a Wess--Zumino--Witten model with
topological coupling $k$ (SU($n$)$_k$ WZW) with central charge
$c=k(n^2-1)/(k+n)$.  
We use periodic boundary conditions (PBCs) rather than hard wall
boundary conditions in our DMRG calculations, as the latter show
even/odd oscillation triggered by the boundaries.  These
would render the interpretation of our results for the category III
models below ambivalent.

\begin{figure}[t]
  \centering
  \vspace{10pt}
  \includegraphics[scale=1.05]{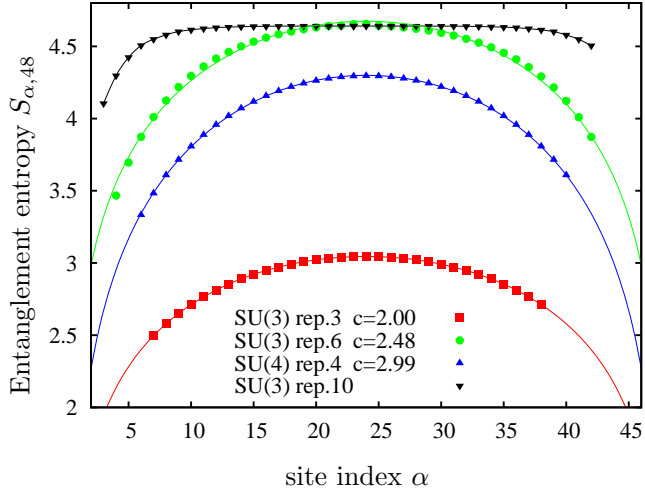}  
    \caption{(Color online) Entanglement entropy (48 sites, PBCs) for
    the critical SU($n$) Heisenberg chains with SU(3) rep.\,$\bs{3}$
    ($N_{\rm cut}\!=\!5000$ kept DMRG states),
    rep.\,$\bs{6}$ ($N_{\rm cut}\!=\!5000$), rep.\,$\bs{10}$ ($N_{\rm
      cut}\!=\!1650$), and SU(4) rep.\,$\bs{4}$ ($N_{\rm
      cut}\!=\!12000$).  Except of SU(3) rep.\,$\bs{10}$, all models
    are gapless and nicely follow the sinusoidal shape according
    to~\eqref{ee-formula}.  The latter clearly develops a plateau
    structure as expected for gapped models.}
  \label{fig:su3su4-critical}
\end{figure}

If the models exhibit a gap in the excitation spectrum, the
entanglement entropy does not have the sinusoidal shape but saturates
after a few sites.  If the models have dimerized (or trimerized,
tetramerized, etc.)  ground states, this manifests itself in the
entanglement entropy by oscillations of the entropy with an
oscillation period of two sites (or three, four, etc.) reflecting that
the ordered states get pinned by numerical noise.  The numerical
results are shown in Fig.\,\ref{fig:su3su4-critical}. 
For the nearest neighbor SU($n$) HMs with fundamental representation
($\lambda=1$), we find the sinusoidal shape\,\eqref{ee-formula} with
fitting parameter $c=n-1$ for $n=2,3,4$\,\cite{fuehringer-08adp922}.
For the SU(3) representation $\bs{6}$ HM, we find a critical model
with a central charge larger than $c=2$, with the difference due to
marginal operator perturbations. For larger chain lengths, the
associated inverse logarithmic corrections disappear and the central
charge approaches $c=2$, as expected.  The situation is hence
comparable to the spin\,$3/2$ HM\,\cite{ziman-87prl140}.  This implies
that the SU(3) representation $\bs{6}$ HM belongs to the SU(3)$_1$ WZW
universality class.  For the ten--dimensional representation of SU(3)
($\lambda=3$), we find that the entanglement entropy saturates after a
few sites. This indicates a gap (see Fig.\,\ref{fig:su3su4-critical}).

\begin{figure}[t]
  \centering
\vspace{6pt}
\includegraphics[scale=1.03]{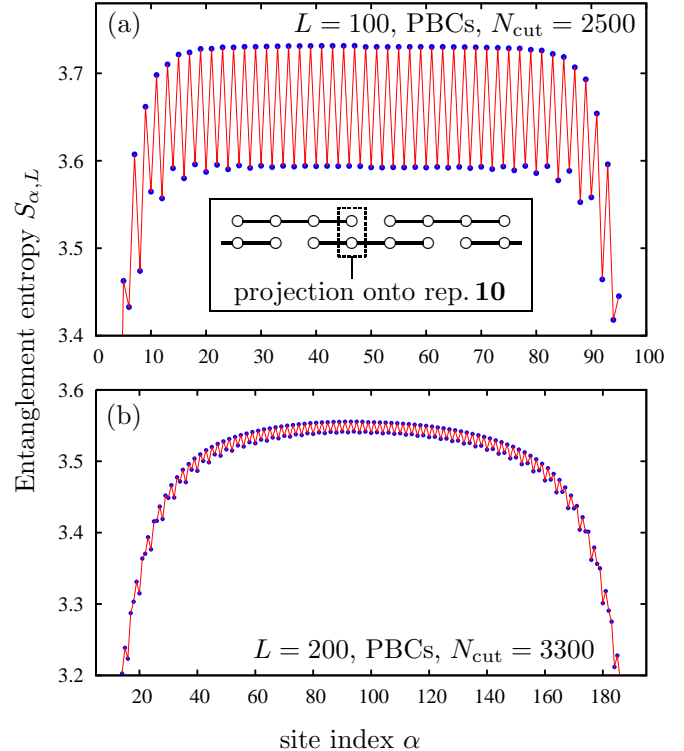}
\caption{Entanglement entropy of the SU(4) rep.\,$\bs{10}$ HM with
  next--nearest neighbor interactions for (a) $J_2=0.5 J_1$ and (b)
  $J_2= 0.15 J_1$, as discussed in the text. The inset in (a)
  illustrates the SU(4) rep.\,$\bs{10}$ VBS state where each circle
  denotes a fundamental SU(4) representation and the horizontal line
  denotes antisymmetric coupling.  Four connected circles hence represent
  an SU(4) singlet.}
  \label{fig:su4r10} 
\end{figure}

{\it Category III models.---}We have further performed extensive
calculations for the HM of the ten--dimensional representation of
SU(4) ($\lambda=2$) with nearest and next--nearest neighbor
interactions.  This is the simplest example belonging to the third
category.  Let us first look at the corresponding VBS state shown in
the inset of Fig.\,\ref{fig:su4r10}a.  The state is two--fold
degenerate and invariant under translations by two lattice spacings.
Its parent Hamiltonian involves three--site
interactions~\cite{greiter-07prb184441}. The correlations decay
exponentially with length $\xi$ (see Table~\ref{tab:corr}).

For nearest neighbor interactions, we find that the model is gapless
and critical as required by the Affleck--Lieb
theorem~\cite{affleck-86lmp57}.  The fitted central charge is larger
than 3 in accordance with the SU(4)$_1$ WZW universality class in the
presence of logarithmic corrections.  Following our categorization,
applying a next--nearest neighbor interaction should force the system
in a gapped phase with confined spinons.  As a first test, we
approximate the Hamiltonian of the VBS state within the $J_1$--$J_2$
model $H_{\rm
  VBS}\approx\sum_i\bs{S}_i\bs{S}_{i+1}+1/2\bs{S}_i\bs{S}_{i+2}$.  The
corresponding entanglement entropy is shown in
Fig.\,\ref{fig:su4r10}a.  It shows a strongly oscillating behavior
indicating dimerization, as expected from the analogy with the VBS
state shown in the inset. 
\begin{table}[t]
\begin{center}
\begin{tabular}{cccc}
\hline\hline
SU($n$)&~~~~~~~~\setlength{\unitlength}{6pt}
\begin{picture}(1,1)(0.3,0.3)
\linethickness{0.3pt}
\multiput(0,0)(1,0){2}{\line(0,1){1}}
\multiput(0,0)(0,1){2}{\line(1,0){1}}
\end{picture}~~~~~~~~&~~~~~~~\setlength{\unitlength}{6pt}
\begin{picture}(2,1)(0.3,0.3)
\linethickness{0.3pt}
\multiput(0,0)(1,0){3}{\line(0,1){1}}
\multiput(0,0)(0,1){2}{\line(1,0){2}}
\end{picture}~~~~~~~~&~~~~~\setlength{\unitlength}{6pt}
\begin{picture}(3,1)(0.3,0.3)
\linethickness{0.3pt}
\multiput(0,0)(1,0){4}{\line(0,1){1}}
\multiput(0,0)(0,1){2}{\line(1,0){3}}
\end{picture}~~~\\[5pt] \hline
$n=2$ & 0.2411 & cat.\,II &
0.33 \\
$n=3$ &0.45& $\geq 0.5^\star$ & cat.\,II\\
$n=4$ & $\approx$ 0.5$^\star$ & ?\\
\hline\hline \\[-10pt]
\end{tabular}
\end{center}
\caption{Critical couplings\,\cite{okamoto-92pla433,roth-98prb9264,ziman-87prl140,corboz-07prb220404} of $J_1$--$J_2$ models of different SU($n$) reps.\ are shown. For representations belonging to the category II with Haldane gap, no phase transition occurs. All other SU($n$) models tend to $n$--merize (\ie dimerize for $n\!=\!2$, etc.)\,\cite{corboz-07prb220404,ziman-87prl140}. We further conjecture, that the critical couplings $(J_2/J_1)_c$ increase when $n$ and/or $\lambda$ is increased. Consequently, the SU(4) rep.$\bs{10}$ chain is expected to tetramerize for a critical coupling which is larger than $(J_2/J_1)_c\approx 0.5$. For the SU(3) rep.$\bs{6}$ and the SU(4) rep.$\bs{4}$ model, only preliminary results are available, as indicated by the asterisks.}
\label{tab:frust}
\vspace{-5pt}
\end{table}%
The conjecture that the HMs behave as the associated VBS models thus
is confirmed for this type of models.  It still remains to be shown, however, that this gap is not due to
a tetramerized phase as expected for a frustrated SU(4) model driven
by $J_2$.  To rule this out, we have collected the critical couplings
of all relevant SU($n$) $J_1$--$J_2$ models in Tab.\,\ref{tab:frust}.
From there, one would expect that a transition into a tetramerized
phase for SU(4) rep.\,$\bs{10}$ would only occur for some critical
coupling $(J_2/J_1)_c>0.5$.  By contrast, we find a dimer phase rather
than a tetramer phase, and for couplings smaller than the expected
critical coupling.  In our DMRG calculations, we observe the existence
of the dimer phase for a value as small as $J_2\!\approx\!0.15 J_1$
(see Fig.\,\ref{fig:su4r10}b).  The oscillation amplitude of the
dimerization becomes weaker but is still present.  Our data cannot be
understood in the context of frustrated HMs, but are fully consistent
with our conjecture of spinon confinement triggered by a next--nearest
neighbor interaction.

Physically, the opening of a gap can be understood as follows.  As we
apply a next-nearest neighbor coupling $J_2$, pairs of neighboring
sites effectively cluster into new sites, and pairs of spins
transforming under the original representation form new
representations on the new sites.  The state hence dimerizes on the
original lattice, while it remains invariant under translations of the
new lattice.  The original representations implied that spinons were
deconfined and no gap occurred.  Spinons in the effective model with
spins transforming under the new representations, however, are
confined and the spectrum is gapped.  The next--nearest neighbor
coupling $J_2$ in our category III model here hence produces an effect
similar to the effect produced by a coupling of two spin 1/2 chains
into a spin ladder.

{\it Conclusion.---}We have shown that a classification of SU($n$)
spin chains with regard to the existence of a Haldane gap obtained
previously for VBS models applies to SU($n$) Heisenberg chains as
well.  The results provide evidence in favor of our hypothesis that
this gap can be interpreted as the zero point energy of an oscillator
describing the relative motion of confined spinons.

{\it Acknowledgement.}---SR acknowledges helpful discussions with
A.\,M.\,L\"auchli and RT with I.\,Affleck at the LXXXIX Les Houches summer school.

\end{document}